\documentstyle[aps,multicol,prl,psfig,epsf]{revtex}
\begin{document}
\title
{
Scaling function
for 
the noisy Burgers equation
in the soliton approximation
}
\author{Hans C. Fogedby}
\address{
\thanks{Permanent address}
Institute of Physics and Astronomy,
University of Aarhus, DK-8000, Aarhus C, Denmark
\\
and
NORDITA, Blegdamsvej 17, DK-2100, Copenhagen {\O}, Denmark
}
\date{\today}
\maketitle
\begin{abstract}
We derive the scaling function for the one
dimensional noisy Burgers equation in the two-soliton 
approximation within the weak noise canonical phase space
approach. The result is in agreement with an 
earlier heuristic expression  and exhibits the correct scaling
properties. The calculation presents the first step
in a many body treatment of the correlations in
the Burgers equation.

\end{abstract}
\pacs{PACS numbers: 05.10.Gg, 05.45.-a, 05.45.Yv}
\begin{multicols}{2}
The strong coupling aspects of driven nonequilibrium systems
present an important challenge in statistical physics.
The phenomena in question are widespread, including
turbulence, interface and growth problems, and chemical
reactions.

Here the noisy Burgers equation or equivalently the
Kardar-Parisi-Zhang (KPZ) equation, describing the growth
of an interface, is one of the simplest models
of a driven system showing scaling and pattern
formation.

In one dimension the noisy Burgers equation for the slope 
$u=\nabla h$ of a growing interface has the form
\cite{Forster}
\begin{eqnarray}
&&\left(\frac{\partial}{\partial t}-\lambda u\nabla\right)u=
\nu\nabla^2u + \nabla\eta ~, 
\label{bur1}
\\
&&\langle\eta(xt)\eta(x't')\rangle = \Delta\delta(x-x')\delta(t-t') ~.
\label{bur2}
\end{eqnarray}
The height $h$ is then governed by the KPZ equation \cite{kpz}
\begin{eqnarray}
\partial h/\partial t = \nu\nabla^2 h + (\lambda/2)(\nabla h)^2 + \eta ~.
\label{kpz}
\end{eqnarray}

In (\ref{bur1}) and (\ref{kpz}) $\nu$ is the damping, 
$\lambda$  the nonlinear
mode coupling, and $\eta$ a Gaussian 
white noise  of strength $\Delta$, correlated according to
(\ref{bur2}). The
equation (\ref{bur1}) is moreover invariant under the Galilean
transformation
\begin{eqnarray}
x\rightarrow x - \lambda u_0t ~~,~~~ u\rightarrow u + u_0 ~.
\label{gal}
\end{eqnarray}
The Burgers equation (\ref{bur1}) and its KPZ equivalent in one and
higher dimensions and related models in the same universality class
have been studied intensely in recent years
owing to their importance as models 
for a class of nonequilibrium
systems
\cite{scal,field,Halpin95}.

We have in a series of papers analyzed the one dimensional
case defined by (\ref{bur1}) and (\ref{bur2}) in an attempt to uncover the
physical mechanisms underlying the pattern formation and scaling behavior
\cite{Fogedby1}.
Emphasizing that the noise strength $\Delta$ is the {\em relevant} 
nonperturbative parameter, driving the system into a stationary state,
the method was initially based on a weak noise saddle point
approximation to the Martin-Siggia-Rose functional formulation
\cite{msr} of the
noisy Burgers equation. 
This work was a continuation of earlier work based on the mapping
of a solid-on-solid model onto a continuum spin model \cite{Fogedby95}.
More recently the functional approach has been
superseded by a {\em canonical phase space method} \cite{Fogedby2}
deriving from the 
symplectic structure \cite{psm} of the Fokker-Planck 
equation associated with the
Burgers equation.

The functional or the equivalent phase space approach valid in the
weak noise limit $\Delta\rightarrow 0$ yields coupled {\em deterministic}
mean field equations
\begin{eqnarray}
\left(\frac{\partial}{\partial t}-\lambda u\nabla\right)u &&=
\nu\nabla^2 u -\nabla^2p ~,
\label{mfe1}
\\
\left(\frac{\partial}{\partial t}-\lambda u\nabla\right)p &&=
-\nu\nabla^2 p ~,
\label{mfe2}
\end{eqnarray}
for the slope $u$ and a canonically conjugate {\em noise field} $p$
(replacing the stochastic noise $\eta$), determining  orbits in a 
canonical phase
space and replacing the stochastic Burgers equation (\ref{bur1}).
The equations (\ref{mfe1}) and (\ref{mfe2}) derive from
a {\em principle of least action} with Hamiltonian density
${\cal H} = p(\nu\nabla^2u + \lambda u\nabla u - (1/2)\nabla^2 p)$
and action $S$ associated with an orbit 
$u'\rightarrow u''$ traversed in time $t$,
\begin{eqnarray}
S(u'',t,u') = \int_{0,u'}^{t,u''}dtdx
\left(p\frac{\partial u}{\partial t} - {\cal H}\right) ~.
\label{act}
\end{eqnarray}

The action is of central importance
and serves as a {\em weight} for the 
nonequilibrium configurations
(cp. the Boltzmann-Gibbs factor $\exp(-E/kT)$ for
equilibrium systems). The action moreover gives access to the
time dependent and stationary probability distributions
\begin{eqnarray}
P(u'',t,u')\propto\exp\left[-S(u'',t,u')/\Delta\right] ~,
\label{dist} 
\end{eqnarray}
and
$
P_{\text{st}}(u'') = \lim_{t\rightarrow\infty}P(u'',t,u') ~,
$
and associated moments, e.g., the slope correlations
\begin{eqnarray}
\langle uu\rangle(xt) = 
\int\prod du~ u''(x)u'(0)P(u'',t,u')P_{\text{st}}(u') ~.
\label{cor}
\end{eqnarray}
The equations (\ref{mfe1}) and (\ref{mfe2}) admit static
soliton solutions
\begin{eqnarray}
u_s^\mu = \mu u\tanh[k_s x] ~, ~~~k_s=\lambda u/2\nu ~, ~~~~\mu=\pm 1 ~,
\label{sol}
\end{eqnarray}
moving solitons are generated by the Galilean boost (\ref{gal}).
Denoting the right and left boundary values by $u_+$ and $u_-$, respectively,
the velocity is given by 
\begin{eqnarray}
u_+ +  u_- = -2v/\lambda ~.
\label{rel}
\end{eqnarray}
The index $\mu$ labels the {\em right hand} soliton
for $\mu = 1$ on the `noiseless' manifold  $p=0$, 
also a solution of the damped noiseless
Burgers equation for $\eta =0$; and the noise-excited {\em left hand}
soliton for $\mu = -1$ on the `noisy' manifold $p=2\nu u$, 
a solution of the growing
(unstable) noiseless Burgers equation for $\nu\rightarrow -\nu$. 
The wavenumber $k_s$ 
sets the inverse length scale.
The field equations also 
admit linear mode solutions superimposed on the solitons;
for $\lambda = 0$ they  become the usual diffusive modes of the
driven equation \cite{Fogedby2}.

The physical picture emerging from this analysis is a 
many body formulation
of the pattern formation of a growing interface in terms of a dilute
gas of propagating solitons with superimposed linear modes. The 
formulation also associates energy $E=\int dx{\cal H}$ and momentum
$\Pi = \int dx u\nabla p$ with a soliton mode, yielding the 
dispersion law 
\begin{eqnarray}
E\propto(\lambda/\nu^{1/2}) \Pi^z ~,
\label{disp}
\end{eqnarray}
with dynamic exponent $z=3/2$ and it follows that the strong coupling
fixed point features are associated with the soliton dynamics, i.e., 
defect or domain wall excitations.

In this Letter we pursue the form of the slope correlations
(\ref{cor}); the basic building block in the many body formulation.
We focus in particular on 
the scaling function which is of central importance.
The dynamic scaling hypothesis
\cite{kpz,Halpin95} and general arguments based on the renormalization
group fixed point structure \cite{scal,Krug97}
imply the following long time-large distance form of the
slope correlations in the stationary state:
\begin{eqnarray}
\langle uu\rangle (xt) = (\Delta/2\nu) x^{-2(1-\zeta)}
G\left(x/\xi(t)\right) ~.
\label{scal1}
\end{eqnarray}
Here $G$ is the scaling function and  $\zeta = 1/2$  the 
roughness exponent inferred from the 
known stationary probability distribution \cite{Huse85}
\begin{eqnarray}
P_{\text{st}}(u)\propto\exp\left[-(\nu/\Delta)\int dx~ u^2\right] ~.
\label{stat2}
\end{eqnarray}
Within the canonical phase space approach (\ref{stat2}) follows from
the structure of the zero-energy manifold which attracts the phase space
orbits for $t\rightarrow\infty$. The dynamical exponent $z=3/2$
then follows from the scaling law $\zeta + z =2$
implied by the Galilean invariance (\ref{gal}) \cite{kpz}.
In the present approach the exponent $z$ is inferred from the (gapless)
soliton dispersion law
(\ref{disp}). 
Finally, the growth of lateral correlations along the interface is
characterized by the time dependent correlation length 
$\xi(t)$. In the nonlinear nonequilibrium 
Burgers case $\xi(t)$ describes the
propagation of solitons and is given by
$\xi(t) = \left(\Delta/\nu\right)^{1/3}(\lambda t)^{2/3}$.
In the linear equilibrium 
Edwards-Wilkinson case \cite{ew} $\xi(t)$ characterizes
the growth of diffusive modes and has the form
$\xi(t) = (\nu t)^{1/2}$.
It moreover follows from the `fluctuation-dissipation theorem' (\ref{stat2})
that $u$ is uncorrelated and that the static 
correlations have the 
form,
$\langle uu \rangle(x) = (\Delta/2\nu)\delta(x)$,
independent of $\lambda$.
This is consistent with $\zeta = 1/2$ and the limiting form of the scaling
function $\lim_{w\rightarrow\infty}G(w) = 1$ for $x\gg\xi(t)$.
In the dynamical regime for
$\xi(t)\gg x$ the correlation decay, i.e., $\langle uu\rangle(x,t)
\rightarrow\langle u\rangle \langle u\rangle = 0$, and the
scaling function vanishes like $G(w)\propto w^{2(1-\zeta)}$
for $w\rightarrow 0$.

Generally, the scaling function can be inferred from the asymptotic
properties of the slope correlations (\ref{cor}). In order to evaluate
those  we must i) determine an orbit from $u'$ to $u''$ in time
$t$ by solving the field equations (\ref{mfe1}) and (\ref{mfe2}) as an
initial-final value problem in $u$ ($p$ is a slaved variable),
ii) evaluate the associated action (\ref{act}) in order to weigh the orbit
and determine
$P$ (note that $P_{\text{st}}$ is given by (\ref{stat2})),
and, finally, iii)
integrate over initial and final configurations $u'$ and $u''$. 
Even in the one dimensional case 
discussed here such a calculation appears
rather formidable in the general multi-soliton - linear mode case and
we must resort to partial results.

In the weak noise limit the action  (\ref{act}) according to
(\ref{dist}) provides a {\em selection criterion} determining the dominant
dynamical configuration contributing to the distribution.
For $\Delta\rightarrow 0$ an important contribution
to the growth
morphology is constituted by  two-soliton configurations or
pair excitations
\begin{eqnarray}
u_2(x,t) = u^{+}_s(x-vt-x_1)+u^{-}_s(x-vt-x_2) ~,
\label{twosol}
\end{eqnarray}
obtained by matching two Galilei-boosted static solitons of opposite
parity ($\mu = \pm 1$) (\ref{sol}) centered at $x_1$ and $x_2$ with
soliton separation $\ell = |x_1 - x_2|$ and amplitude $2u$. 
According to the soliton condition (\ref{rel}) the pair excitation 
propagates with velocity $v=-\lambda u$ and has
vanishing slope field  $u=0$ at the boundaries,
corresponding to a horizontal interface.

Whereas the solitons (\ref{sol}) for $\mu=\pm 1$ lie on the transient
and stationary submanifolds (separatrixes) for $p_s=0$ (the `noiseless' kink)
and $p_s=2\nu u_s$ (the `noisy' kink), respectively, and constitute
the `quarks' in the many body formulation, the pair excitation
(\ref{twosol}), satisfying the boundary conditions, is the elementary
excitation or `quasi particle' (in the Landau sense) in the present
scheme and is characterized by the 
dispersion law (\ref{disp}). The pair excitation is an approximate
solution to the field equations (\ref{mfe1}) and (\ref{mfe2}) with
a finite lifetime \cite{Fogedby01a}. 
Over a time scale controlled by the damping $\nu$
the pair decays into
diffusive modes; this is consistent with the observation that the
phase space orbits approach the zero-energy manifold for 
$t\rightarrow\infty$.

\begin{figure}
{
\centerline
{
\epsfxsize=13cm
\epsfbox{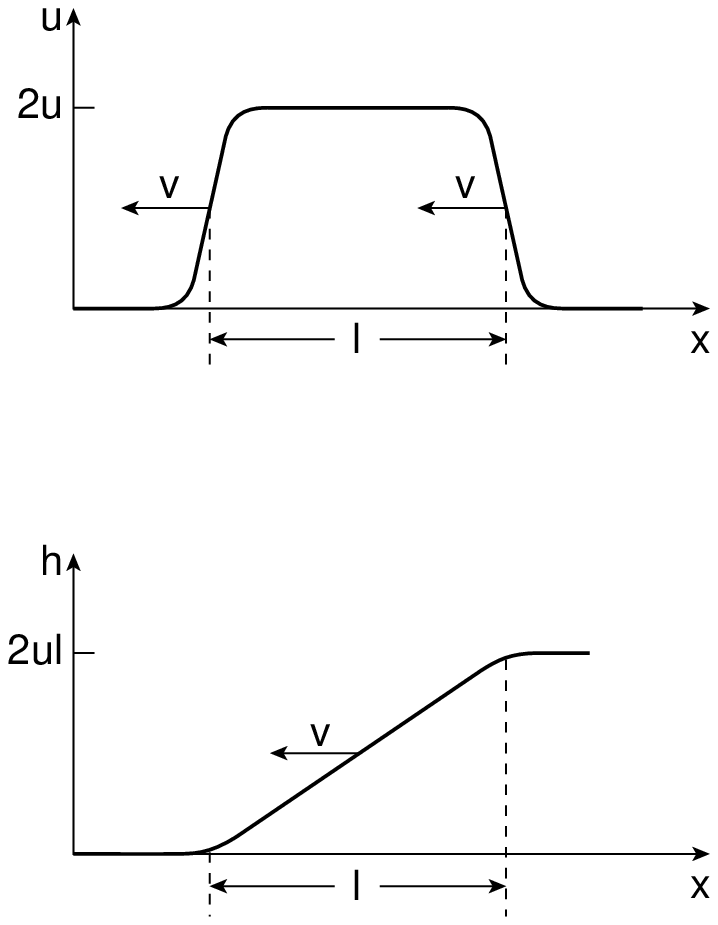}
}
}
\caption{
The two-soliton growth mode in the slope field $u$  and
the associated height field $h$, $u=\nabla h$. The
pair soliton excitation has amplitude $2u$ and size
$\ell$.
}
\label{fig1}
\end{figure}

Unlike a general multi-soliton configuration which changes in
time owing to soliton-soliton collisions, the pair excitation
preserves its shape over a
finite time period, see ref. \cite{Fogedby2,Fogedby01a}. Imposing
periodic boundary conditions for the slope field the motion of a pair
with amplitude $2u$ corresponds to a simple growth mode where the
height field $h$, i.e., the integrated slope field, increases 
layer by layer for each revolution of the soliton pair in a system of 
size $L$. 
From the KPZ equation (\ref{kpz})
it follows that 
$\langle dh/dt\rangle = (\lambda/2)\langle u^2\rangle$ in a stationary
state. Setting $u\rightarrow 2u$ this is consistent with the increase
$\Delta h = 2u\ell$ during the passage time 
$\Delta t = \ell/v = \ell/\lambda u$ for a soliton pair of size $\ell$. 
In Fig.~\ref{fig1} we have depicted the two-soliton growth mode in the slope
field $u$ and the associated height field $h$.
The pair excitation, which can also be conceived as a bound state
composed of two solitons, has the amplitude $2u$, size $\ell$,
carries energy $E_2=-(16/3)\nu\lambda|u|^3$, momentum
$\Pi_2 = -4\nu u|u|$, and action
\begin{eqnarray}
S_2 = \frac{4}{3}\nu\lambda|u|^3 t ~.
\label{act2}
\end{eqnarray}

Using the definition (\ref{cor}) it is an easy task to evaluate 
the contribution to the
slope correlations from a single pair.
The normalized stationary distribution $P_{\text{st}}$ is obtained
from (\ref{stat2}) by insertion of (\ref{twosol}). Considering  the inviscid
limit for $\nu\rightarrow 0$ we have
\begin{eqnarray}
&&P_{\text{st}}(u,\ell) = \Omega^{-1}_{\text{st}}
\exp\left[-4(\nu/\Delta)u^2\ell\right] ~,
\label{stat3}
\\
&&\Omega_{\text{st}} = (\pi\Delta/\nu)^{1/2}L^{3/2} ~.
\label{norm1}
\end{eqnarray}
Correspondingly, inserting (\ref{act2}) in (\ref{dist}) the normalized
soliton pair transition probability is 
\begin{eqnarray}
&&P_{\text{sol}}(u,t) = \Omega^{-1}_{\text{sol}}
\exp\left[-(4/3)(\nu/\Delta)\lambda |u|^3t\right] ~,
\label{dsol}
\\
&&\Omega_{\text{sol}} = (2/3)\Gamma(1/3)
[(3/4)(\Delta/\nu)(1/\lambda t)]^{1/3} ~.
\label{norm2}
\end{eqnarray}
We note that the normalization factor $\Omega_{\text{st}}$ for the stationary
distribution varies as $L^{3/2}$ and that the distribution thus
vanishes in the infinite size limit; moreover, the mean size of a pair
is equal to $L$, characteristic of an extended excitation 
(a string).
Likewise the transition probability $P_{\text{sol}}$ goes to zero 
for large times in accordance with the decay of a soliton pair
into diffusive modes.

\begin{figure}
{
\centerline
{
\epsfxsize=13cm
\epsfbox{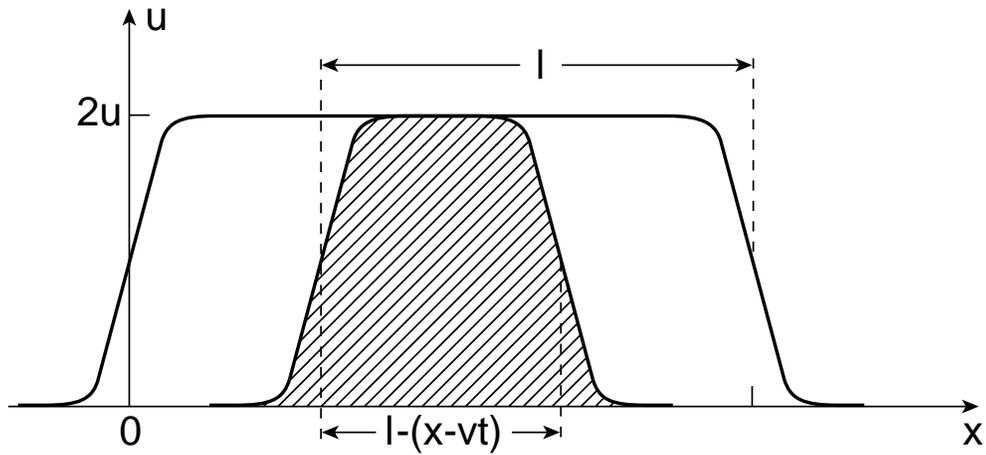}
}
}
\caption{
The two-soliton configuration of size $\ell$ and amplitude
$2u$. The shaded area of size $2\ell - x$ yields a contribution
to the slope correlation function.
}
\label{fig2}
\end{figure}

The evolution of $\langle uu\rangle(xt)$ in the two-soliton sector
is straightforward. The final configuration $u''$ is simply the
initial configuration $u'$ displaced $vt$ along the axis with
no change of shape, i. e., $u''(x)=u'(x+vt)$, $v=-\lambda u'$. 
Noting that the integral over $u'$ and $u''=u'$ only contributes 
when the pair configurations overlap and integrating over the size
$\ell$ we obtain  the slope correlations
\begin{eqnarray}
\langle uu\rangle(xt) =
\frac{\ell_0}{L}
\frac
{
\int du
e^{-\frac{4}{3}|u|^3\frac{t}{t_s}}
e^{-4u^2|\frac{x}{L}+u\frac{t}{t_s}|}
C_1(u)
}
{
\int du
e^{-\frac{4}{3}|u|^3\frac{t}{t_s}}
C_2(u) 
}
~,
\label{cor2}
\end{eqnarray}
where the cut-off functions originating
from the overlap are given by
$C_1(u)=1/4u^2-(1+1/4u^2)\exp(-4u^2)$
and
$C_2(u) = (1/4u^2)(1-\exp(-4u^2))$, respectively.
In order to facilitate the discussion of (\ref{cor2}) we
have introduced the noise-induced length and time scales
$\ell_0=\Delta/\nu$
and $t_0= \Delta/\nu\lambda$; note that $\lambda=\ell_0/t_0$,
and, moreover, the crossover or saturation time
$t_s=t_0(L/\ell_0)^{3/2}$; the correlation length is then
$\xi=\ell_0(t/t_0)^{2/3}$. 
The expression (\ref{cor2}) holds for $t>0$ and is even in $x$ 
(seen by changing $u$ to $-u$). It samples the soliton pair propagating 
with velocity $\lambda u$ and is in general agreement with
spectral form discussed in the `quantum' treatment in 
\cite{Fogedby1}. 
In Fig.~\ref{fig2} we have shown the two-soliton overlap configurations
contributing to the slope correlations.

The weight of single soliton pair is of order $1/L$ and the correlation
function $\langle uu\rangle$ thus vanishes in the thermodynamic limit
$L\rightarrow\infty$. For a finite system $L$ enters 
setting a length scale together with the saturation time 
$t_s\propto L^{3/2}$
defining a time scale, and $\langle uu\rangle$ is a function of
$x/L$ and $t/t_s$ as is the case for the two-soliton expression
(\ref{cor2}). This dependence should be compared with the wavenumber 
decomposition of $\langle uu\rangle$ for $\lambda=0$.
Here $\langle uu\rangle(xt)\propto
(1/L)\sum_{n\neq 0}\exp(-(2\pi n)^2t/L^2)\exp(i\pi nx/L)$,
depending on $x/L$ and $t/L^2$, corresponding to the saturation time
$t_s\propto L^2$, $z=2$. Keeping only one mode for $n=1$ 
$\langle uu\rangle$ has the same structure as in the soliton case. In
the linear case we can, of course, sum over the totality of modes
and in the thermodynamic limit $L\rightarrow\infty$ replace
$(1/L)\sum_n$
by $\int dk/2\pi$ obtaining the intensive correlations
$\langle uu\rangle(xt)=(\Delta/2\nu)(4\pi\nu t)^{-1/2}\exp(-x^2/2\nu
t)$.
Similarly, we expect the inclusion of multi-soliton modes to allow the 
thermodynamic limit to be carried out yielding an intensive correlation
function in the Burgers case.

For a finite system we have in general
\cite{Krug92}
$
\langle uu\rangle(xt)=(1/L)G_L(x/L,t/L^{3/2})
$
with scaling limits: $G_L(x/L,0)\propto\text{const.}$ for $x\sim L$,
$G_L(x/L,0)\propto L/x$ for $x\ll L$ and
$G_L(0,t/L^{3/2})\propto\text{const.}$ for $t\gg L^{3/2}$,
$G_L(0,t/L^{3/2})\propto L/t^{2/3}$ for $t\ll L^{3/2}$.
For $L\rightarrow\infty$ we obtain
$G_L(x/L,t/L^{3/2})\rightarrow (L/x)G(x/t^{2/3})$ in conformity
with (\ref{scal1}). 

It is an important feature of the two-soliton expression (\ref{cor2})
that the dynamical soliton interpretation directly implies 
the correct dependence
on the scaling variables $x/L$ and $t/t_s\propto t/L^{3/2}$, i.e., 
independent of a renormalization group argument. However,
the scaling limits are at variance with $G_L$. Setting, according
to (\ref{cor2}) $\langle uu\rangle(xt)=(\ell_0/L)F(x/L,t/t_s)$,
$F(x/L,0)$ assumes the value $.47$ for $x\ll L$ and decreases
monotonically to the value $\sim .08$ for $x\sim L$, whereas $G_L$
diverges as $L/x$ for $x\ll L$. Likewise, $F(0,t/t_s)$ decays
from $.47$ for $t\ll t_s\propto L^{3/2}$ to $0$ for 
$t\gg t_s$; for $t\sim t_s$ we have $F\sim .15$, 
whereas  $G_L$ diverges as $L/t^{2/3}$ for $t\ll t_s$. 

This discrepancy from the scaling limits is a feature
the two-soliton contribution which only
samples the correlation from a single soliton pair. Moreover, at long times
the soliton contribution vanishes and the scaling function is determined
by the diffusive mode contribution in accordance with the convergence
of the phase space orbits to the stationary zero-energy manifold. We note,
however, the general trend towards a divergence for small values
of $x$ and $t$ is a feature of $F$.

\begin{figure}
{
\centerline
{
\epsfxsize=13cm
\epsfbox{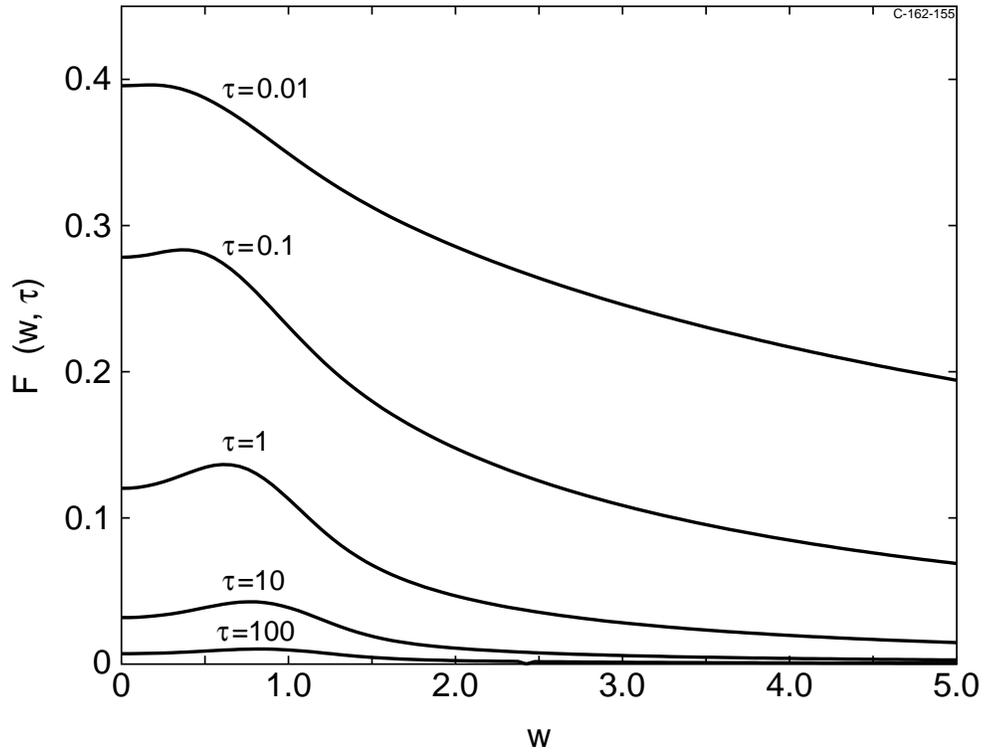}
}
}
\caption
{
Plot of the scaling function $F(w,\tau)$ as a function
of the scaling variable $w=x/\xi\propto x/t^{2/3}$
for a range of values of $\tau=t/t_s\propto t/L^{3/2}$.
}
\label{fig3}
\end{figure}

Introducing the scaling variables $w=x/\xi\propto x/t^{2/3}$ and
$\tau=t/t_s\propto t/L^{3/2}$ we can also express (\ref{cor2})
in the form 
\begin{eqnarray}
\langle uu\rangle(xt)=(\ell_0/L)F_2(w,\tau)~,
\end{eqnarray}
where the scaling function $F$ is now given by
\begin{eqnarray}
F(w,\tau)=
\frac
{
\int du
e^{-\frac{4}{3}|u|^3\tau}
e^{-4u^2|w\tau^{2/3}+u\tau|}
C_1(u)
}
{
\int du
e^{-\frac{4}{3}|u|^3\tau}
C_2(u) 
}
~,
\label{cor3}
\end{eqnarray}
and summarize our findings in 
Fig. \ref{fig3} where we have depicted $F(w,\tau)$ for a
range of $\tau$ values. 
For fixed small $w=x/\xi\propto x/t^{2/3}$ we have
$F\rightarrow .47$ for $\tau=t/t_s\propto t/L^{3/2}\rightarrow 0$;
for large $\tau$ we obtain  $F\rightarrow 0$. 
The motion of the weak maximum towards
smaller values of $w$ for decreasing $\tau$ is a feature of
the functional form of $F$ in (\ref{cor3}), i.e., 
the soliton approximation, and probably not a property of the
true scaling function which 
is not expected to have any particularly distinct features 
\cite{scal,field,Halpin95}.
 
In this Letter we have presented the two-soliton contribution to
the slope correlations and ensuing scaling function within the
weak noise canonical phase space
approach to the noisy Burgers equation in one dimension.
The expression is in accordance
with a general spectral form proposed earlier on the basis of
the many body interpretation of a growing interface and has the
correct scaling dependence. This calculation presents the first
step in a many body or field theoretical treatment of
the correlations in the noisy Burgers equation based on a
transparent physical quasi particle picture of the growth mechanisms and
ensuing morphology. Details will be presented elsewhere.

Discussions with A. Svane, J. Hertz, B. Derrida, M. L\"{a}ssig,
J. Krug and G. Sch\"{u}tz
are gratefully acknowledged.

\end{multicols}
\newpage
\newpage
\end{document}